\documentclass{iauc}
\usepackage{graphicx}
\newcommand\farcs{\mbox{$.\!\!^{\prime\prime}$}}%

\title[Evolution of Field Dwarf Galaxies with GEMS]
{Evolution of Field Dwarf Galaxies with GEMS}

\author[Barazza \& Jogee]   
{Fabio D. Barazza$^1$%
  \thanks{Present address: Department of Astronomy, University of Texas at
    Austin, 1 University Station C1400, Austin, TX 78712-0259, USA},
  Shardha Jogee$^2$} 

\affiliation{$^1$Space Telescope Science Institute, 3700 San Martin Drive,
Baltimore, MD 21218, USA \break 
email: barazza@astro.as.utexas.edu\\[\affilskip]
$^2$Department of Astronomy, University of Texas at Austin, 1 University
Station C1400, Austin, TX 78712-0259, USA}

\pubyear{2005}
\volume{198} 
\pagerange{1--4}
\date{?? and in revised form ??}
\jname{Near-Field Cosmology with Dwarf Elliptical Galaxies}
\editors{Helmut Jerjen, Bruno Binggeli, eds.}
\begin{document}

\maketitle

\begin{abstract}
We present a study of the evolution of late-type field dwarfs over the last
$\sim 1.9$ Gyr, based on $HST$ ACS observations carried out as part of the GEMS
survey. This study is amongst the first to probe the evolution of dwarfs over
such a large timescale. The comparison of structural properties, particularly
size and scale length, indicates that the dwarfs in the redshift range $z \sim$
0.01 to 0.15 (look-back times up to 1.9 Gyr) are more extended than local
dwarfs. We argue that this difference is due to the star formation activity
becoming more centrally concentrated in late-type dwarfs over the last $\sim
1.9$ Gyr. We discuss several possible causes for this evolution. We also find
a lack of clue compact dwarfs in the GEMS sample and interpret this as
indicative of the fact that strong, centrally concentrated star formation is a
feature of evolved dwarfs that are entering their final stages of evolution.
\end{abstract}

\firstsection 
\section{Introduction}
The study of dwarf galaxies outside of galaxy clusters has, so far, mainly been
restricted to the Local Group and the local volume ($< 10$ Mpc). In these
regimes, the dwarf population is dominated by late-type dwarfs (henceforth
referred to as dwarf irregulars, dIs), many of which are isolated (i.e. not
associated with a giant galaxy). On the other hand, early-type dwarfs
(henceforth referred to as dwarf spheroidals, dSphs) are almost absent outside
of groups, but they are frequently satellites of giants and therefore quite
numerous in groups. Except this morphology-density relation no structural
differences have been found between group and field dwarfs. It is therefore
still unclear, for instance, whether dIs are the progenitors of dSphs and what
processes might lead to their transformation. Detailed studies of field dIs
indicate that they generally have an older stellar population and that they
would be able to produce stars for another Hubble time, if they would maintain
their current low star formation rates (\cite{van01,gre04}). Hence, if no
environmental interaction occurs, these objects will not end up as dSphs by
passive evolution.

However, the samples of dwarf galaxies studied so far are, in general, still
too small in order to draw firm conclusions. It is therefore desirable to
enlarge the database not only at $z \sim 0$, which might be realized with the
Sloan Digital Sky Survey, but also for earlier epochs, where signatures of a
possible evolution might be found. In this paper, we present first results
based on a study of dwarfs at intermediate redshifts (up to $z=0.15$,
corresponding to look-back times of up to $\sim$ 1.9 Gyr)\footnote[2]{We assume
in this  paper a flat cosmology with $\Omega_M = 1 - \Omega_{\Lambda} = 0.3$
and $H_{\rm 0}$=65~km~s$^{-1}$~Mpc$^{-1}$.}, observed as part of the Galaxy
Evolution from Morphology and SEDs (GEMS, \cite{rix04}) survey. GEMS consists
of the largest-area multi-filter survey carried out to date with the Advanced
Camera for Surveys (ACS) on the $Hubble Space Telescope$ ($HST$), and might,
therefore, offer the best premise for the study of dwarfs at higher redshifts.
The GEMS survey produced high-resolution ($0 \farcs 05$ correspond to 135 pc at
$z=0.15$) images for $\sim$ 8300 galaxies out to $z \sim$ 1.2, for which
accurate redshifts ($\delta_z / (1+z) \sim 0.02$ down to $R_{Vega} = 24$) are
available from the COMBO-17 project (\cite[Wolf et al. 2004]{wol04}).

\begin{figure}
  \includegraphics[height=70mm,width=70mm]{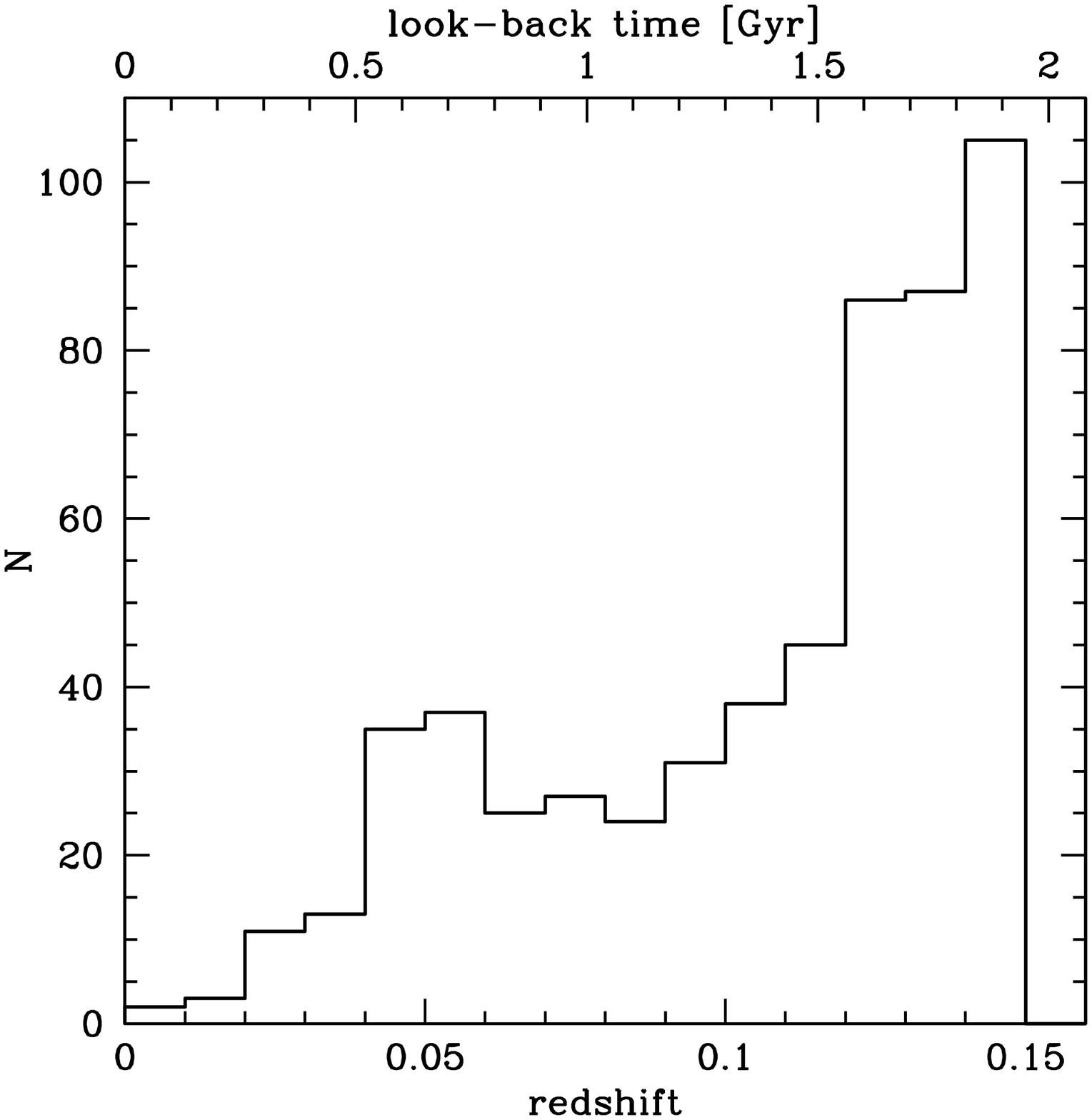}
  \includegraphics[height=70mm,width=70mm]{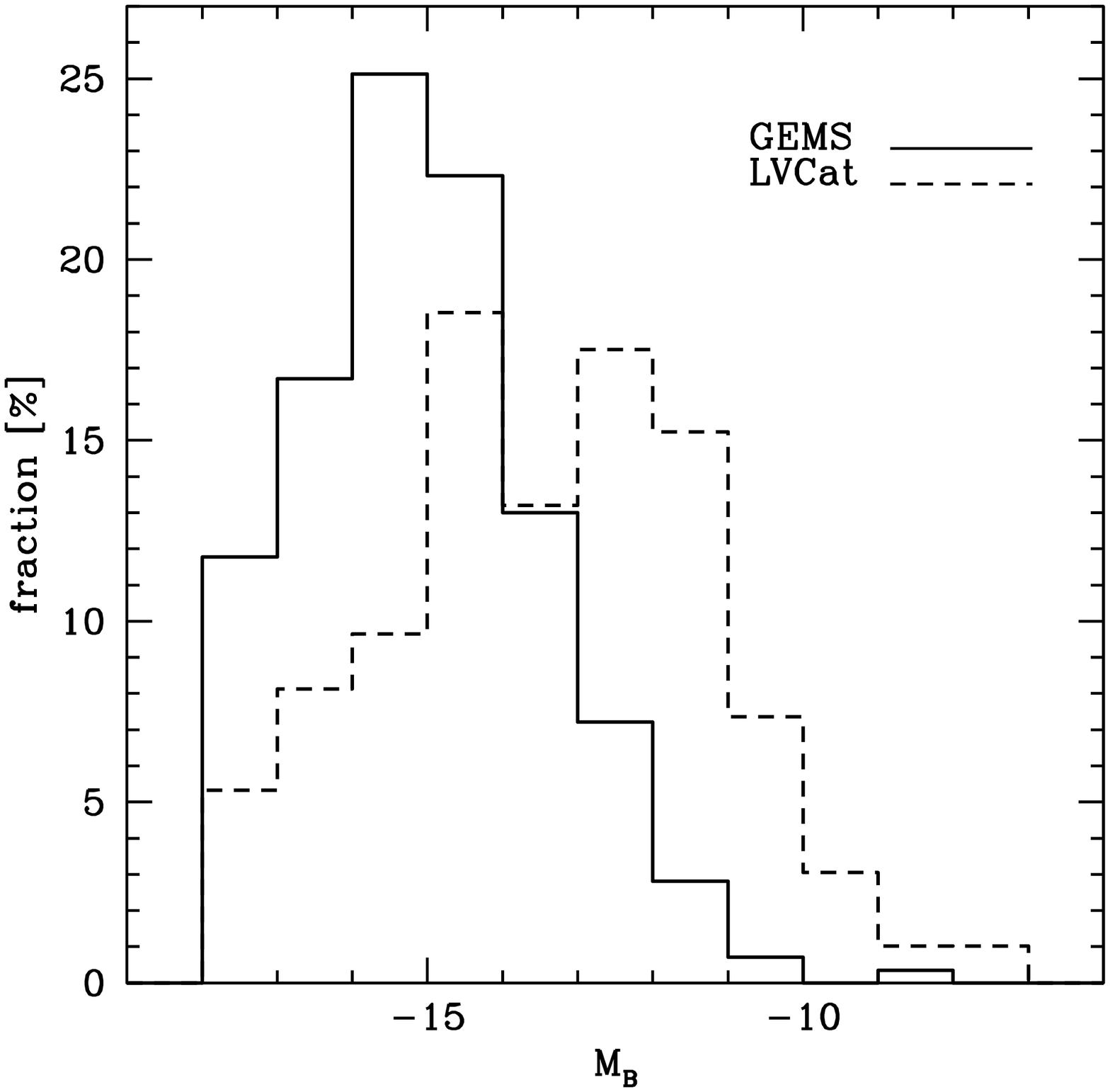}
  \caption{{\it left: } Redshift distribution of the final GEMS dwarf sample.
  {\it right: } Luminosity distributions of the GEMS sample and the catalog of
neighboring galaxies (LVCat).}
\end{figure}

\section{The GEMS sample and the local comparison samples}
The largest local dwarf sample is the catalog of neighboring galaxies
(\cite[Karachentsev et al. 2004]{kar04}), which exhibits a peak around $M_B
\sim -14$ mag in the luminosity distribution of galaxies. We use this catalog
for the comparison of the luminosity distribution of the GEMS sample. However,
since this catalog does not provide additional structural parameters, we
compiled a second sample of local galaxies from the literature, using the
following three sources: \cite[Parodi, Barazza, \& Binggeli (2002, and
references therein)]{par02}, a sample of 80 field and group late-type dwarfs
(LV); \cite{van00}, a sample of 43 isolated dIs (ISO); \cite{cai01} a sample of
15 blue compact dwarfs (BCD). These samples have been selected more or less
randomly. But together they comprise all classes of local dwarfs commonly
studied and they cover all environments outside of clusters. and are,
therefore, appropriate to compare to the GEMS sample.

In order to be complete for galaxies with $M_B < -14$ mag, we had
to restrict the GEMS sample to $z < 0.15$. This limit was determined by
considering the detection limit of the COMBO-17 survey, i.e. at a redshift of
$z \sim 0.15$ the faintest galaxies observed have $M_B \sim -14$ mag. (Note
that we did not apply a $K$-correction, as these corrections for the redshift
range used would be too small to significantly affect our results.) For all
galaxies we determined the surface brightness profile and fitted an exponential
model to it. Based on a rough visual inspection, we excluded all objects with
signs of interaction or overlapping with other objects. The left panel of
Figure 1 shows the redshift distribution of the dwarf sample. More than $70\%$
of the objects are at look-back times of $> 1$ Gyr. The right panel of Figure 1
shows the luminosity distributions of the GEMS sample and the catalog of
neighboring galaxies. The agreement is quite good (keeping in mind that the
GEMS sample is roughly complete for objects with $M_B < -14$ mag), which means
that our study targets systems of similar luminosity across the GEMS and local
samples.

\section{Results}
\begin{figure}
  \includegraphics[height=70mm,width=70mm]{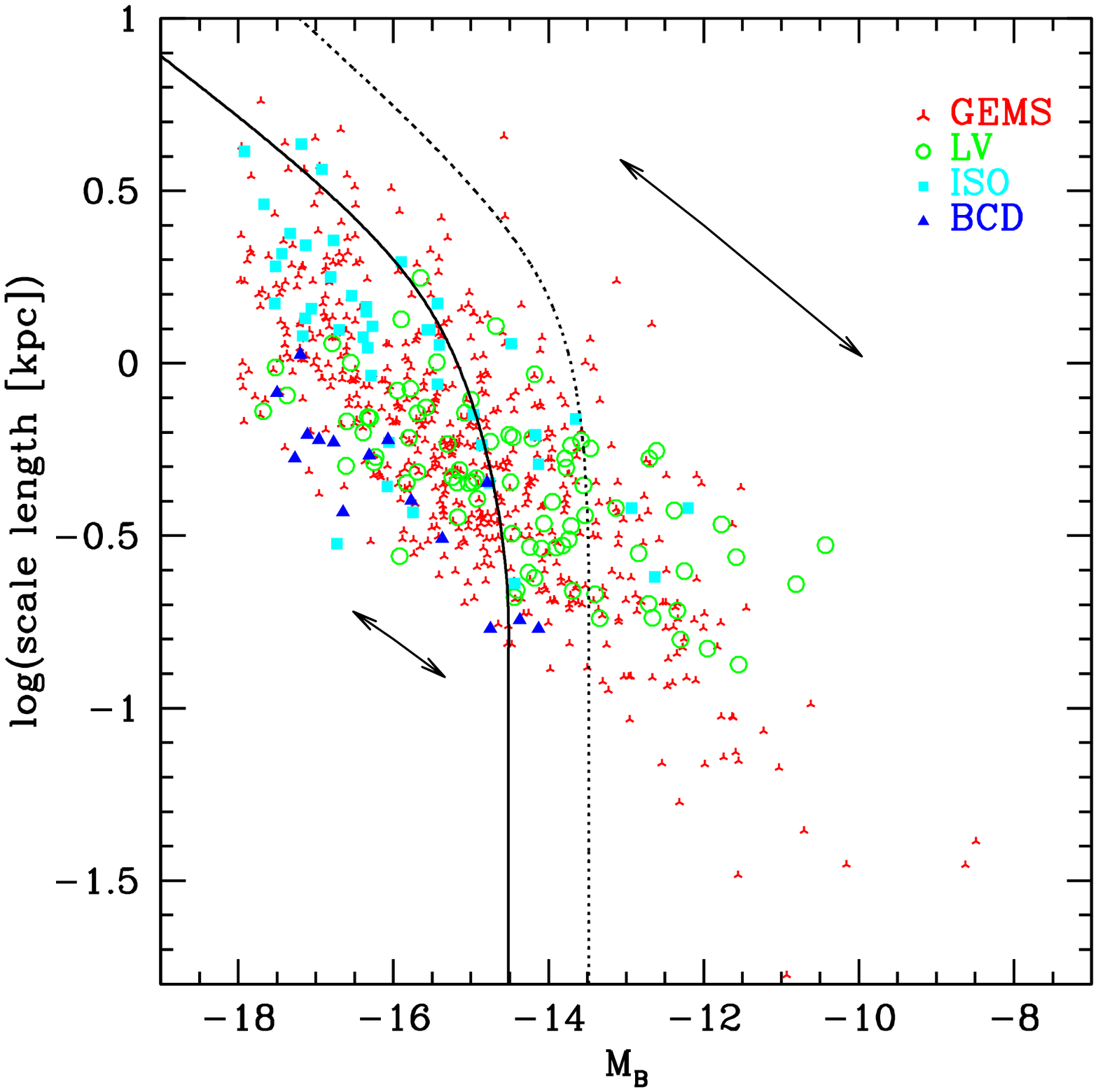}
  \includegraphics[height=70mm,width=70mm]{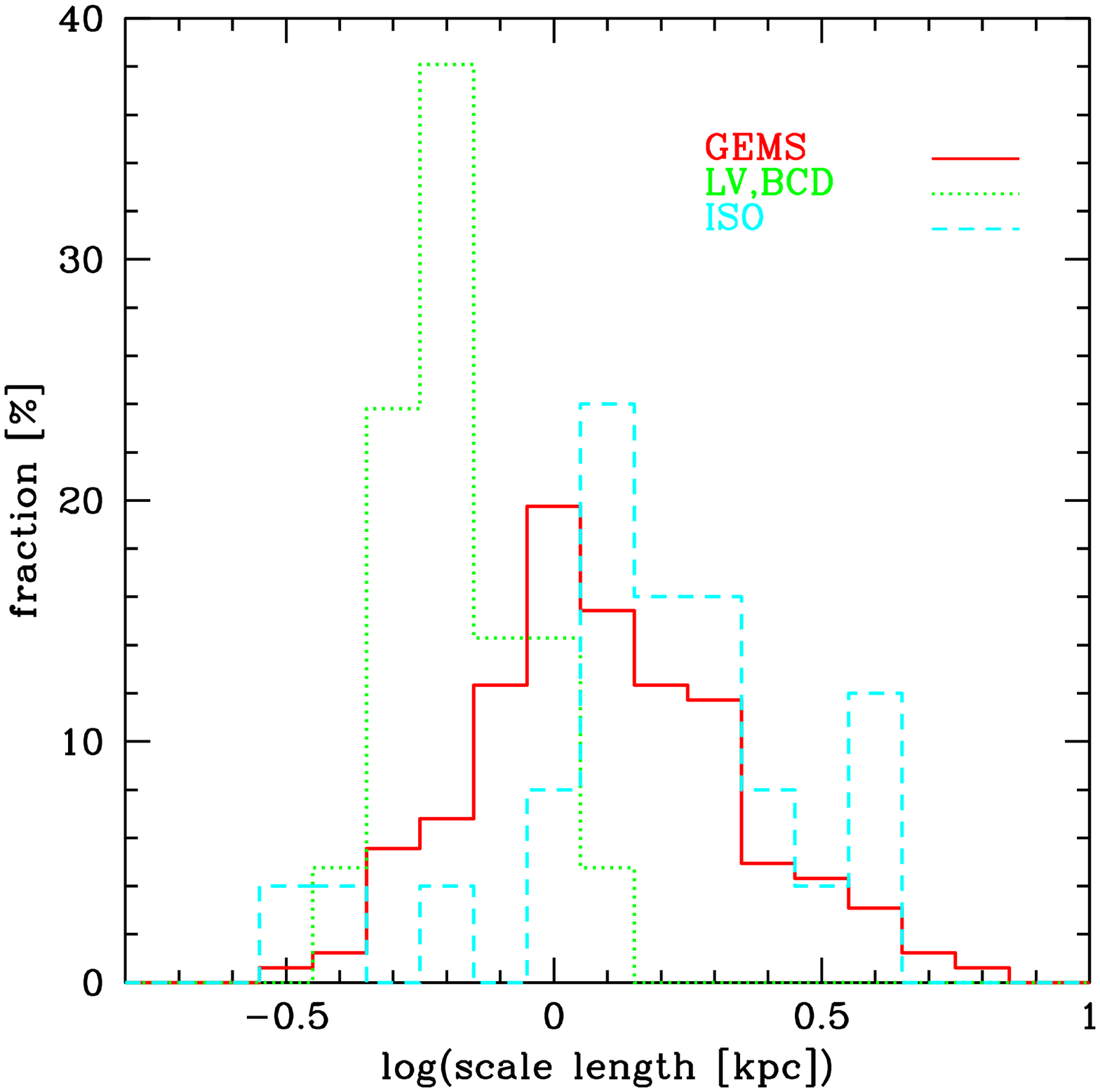}
  \caption{{\it left: } Exponential scale length versus the absolute $B$-band
magnitude. The solid and dashed lines are the $100 \%$ and $50 \%$ completeness
limits, respectively, of the GEMS sample. The bold line with arrows represent
the typical error bars caused by redshift errors and show the direction along
which such errors would move the data points. Note that the measurements for
the GEMS dwarfs have been performed on the $V$-band image, which corresponds to
a rest-frame band somewhere between $B$ and $V$ depending on the redshift.
{\it right: } Histograms of the scale lengths for objects with $M_B < -16$
mag.}
\end{figure}

We compared structural parameters for the local and GEMS dwarfs. The
measurements for the local dwarfs have been performed on $B$-band images,
whereas for GEMS the $V$-band images have been used, which correspond to a
rest-frame band somewhere between $B$ and $V$ depending on the redshift.
Figure 2 shows the main results:
\begin{description}
\item[a) ] GEMS dwarfs with $M_B < -16$ mag in the redshift range 0.01 to 0.15
seem to be more extended than the local dwarfs, particularly the LV and BCD
dwarfs, which tend to be located in galaxy groups or are even satellites of
giant galaxies.
\item[b) ] No significant difference is seen between GEMS dwarfs and the ISO
sample of more isolated local dwarfs.
\end {description}
The interpretation of these results is not straightforward, due to the somewhat
heterogeneous way in which the local samples were selected. In addition, they
also cover a small volume and may, therefore, suffer from cosmic variance. In a
future paper (\cite{bar05}) we will compare our GEMS sample to a sample of
local dwarfs drawn from the SDSS, which covers a large representative volume of
the Universe. We nonetheless discuss here a few possible interpretations of the
results. If we make the common assumption that the GEMS galaxies are primarily
isolated galaxies then finding (b) suggests that there is no significant change
in the properties of isolated dwarfs over the last $\sim 1.9$ Gyr. The more
extended scale lengths in the GEMS galaxies, compared to the local LV and BCD
dwarfs, as reported in (a) above, can be interpreted as resulting from
differences in the spatial extent of star formation sites. Since the LV and BCD
samples are more representative of dwarfs in groups, their less extended blue
scale lengths/star formation might be indicative of gas-removal processes (e.g.
ram pressure stripping) operating in groups. This interpretation receives
support from the fact that a majority of local dwarfs exhibit red color
gradients, i.e. they become redder with increasing radius. In addition, in all
Local Group dwarfs, for which this information is known, the older stellar
population is more extended than the younger one (\cite[van den Bergh
1999]{ber99}). The most likely explanation for these findings is that the star
formation activity must have been more extended in these dwarfs in the past. We
note that passive aging over $1.6$ Gyr, which is the average look-back time of
the GEMS sample, causes the $B-V$ color of a stellar population to redden by
$\sim 0.7$ ($Starburst 99$, \cite{lei99}). This is enough in order to account
for a significant change in the blue scale length.

Another interesting result is the apparent lack of BCDs in the GEMS sample.
Numerous studies indicate that BCDs exist in both isolated and group
environments. We therefore do not assign the absence of BCDs in GEMS as being
caused by environmental differences. Instead, we suggest that the difference is
a temporal one, in the sense that BCD-like structures are a feature of evolved
dwarfs, entering their final stages of evolution.

\begin{acknowledgments}
F.D.B. and S.J. acknowledge support from the National Aeronautics and Space
Administration (NASA) under LTSA Grant NAG5-13063 issued through the Office of
Space Science. Support for GEMS was provided by NASA through GO-9500 from the
Space Telescope Science Institute, STScI, which is operated by the Association
of Universities for Research in Astronomy, Inc. AURA, Inc., for NASA, under
NAS5-26555.
\end{acknowledgments}


\begin{thebibliography}{}
\bibitem[Barazza et al. 2005, in prep.]{bar05} Barazza, F.D. et al. 2005, in
prep.
\bibitem[Cair{\' o}s et al.\ (2001)]{cai01} {Cair{\' o}s, L.~M., Caon, N.,
V{\'{\i}}lchez, J.~M., Gonz{\' a}lez-P{\' e}rez, J.~N., \& Mu{\~ n}oz-Tu{\~
n}{\' o}n, C.} 2001, \textit{ApJS,} 136, 393
\bibitem[Grebel\ 2004]{gre04} Grebel, E.~K.\ 2004, Origin and Evolution of the
Elements, 237 
\bibitem[Karachentsev, Karachentseva, Huchtmeier, \& Makarov\ (2004)]{kar04}
Karachentsev, I.~D., Karachentseva, V.~E., Huchtmeier, W.~K., \& Makarov,
D.~I.\ 2004, \textit{AJ}, 127, 2031 
\bibitem[Letherer et al. 1999]{lei99} Leitherer, C., et al. 1999,
\textit{ApJS}, 123, 3
\bibitem[Parodi, Barazza, \& Binggeli\ (2002)]{par02} {Parodi, B.~R., Barazza,
F.~D., \& Binggeli, B.} 2002, \textit{A\&A,} 388, 29
\bibitem[Rix et al.\ (2004)]{rix04} Rix, H., et al.\ 2004, \textit{ApJS}, 152,
163 
\bibitem[van den Bergh\ (1999)]{ber99} van den Bergh, S.\ 1999, \textit{A\&AR},
9, 273 
\bibitem[van Zee\ (2000)]{van00} {van Zee, L} 2000, \textit{AJ,} 119, 2757 
\bibitem[van Zee\ 2001]{van01} van Zee, L.\ 2001, \textit{AJ}, 121, 2003 
\bibitem[Wolf et al.\ (2004)]{wol04} Wolf, C., et al.\ 2004, \textit{A\&A},
421, 913 
\end{thebibliography}
\end{document}